\documentclass{aa} 
\DeclareUnicodeCharacter{0301}{\'{e}}
\DeclareUnicodeCharacter{02BC}{\'{e}}
\usepackage[hidelinks,colorlinks=true,linkcolor=blue,citecolor=blue]{hyperref}
\usepackage{txfonts}

\usepackage{float}
\usepackage{multirow}
\usepackage{amsmath}
\usepackage{natbib}
\bibpunct{(}{)}{;}{a}{}{,}
\usepackage{graphicx}

\newcommand{\al}{$\alpha$}

\newcommand{\lam}{$\lambda$}

\usepackage{float}
\begin{document} 

\title{The miniJPAS Survey: Detection of the double-core Ly\al~morphology for two high-redshift QSOs}
   \author{P. T. Rahna \inst{1} \fnmsep\thanks{7rehanrenzin@gmail.com}
          \and
          Z.-Y. Zheng \inst{1}
          \fnmsep\thanks{zhengzy@shao.ac.cn}
          \and
          A. L. Chies-Santos \inst{2,1}
          \and
          Z. Cai \inst{3}
          \and
          D. Spinoso \inst{4}
          \and
          I. Marquez \inst{5}
          \and
          R. Overzier \inst{6}
          \and
          L. R. Abramo \inst{7}
          \and
          S. Bonoli \inst{4}
          \and
          C. Kehrig \inst{5}
          \and
        L. A. D\'iaz-Garc\'ia \inst{5}
        \and
        M. Pović \inst{8,5}
        \and
        R. Soria \inst{9,10}
        \and
        J. M. Diego \inst{11}
        \and
        T. Broadhurst \inst{12,13,14}
        \and
        R. M. González Delgado \inst{5}
        \and
        J. Alcaniz\inst{6}
        \and
        N. Ben\'itez\inst{5,14}
        \and
        S. Carneiro\inst{15} 
        \and
        A. J. Cenarro\inst{16}
        \and
        D. Cristóbal-Hornillos\inst{16}
        \and
        R. A. Dupke\inst{6,17,18}
        \and
        A. Ederoclite\inst{16}
        \and
        A. Hernán-Caballero \inst{16}
        \and
        C. L\'opez-Sanjuan\inst{16}
        \and
        A. Mar\'in-Franch\inst{16}
        \and
        C. Mendes de Oliveira\inst{19}
        \and
        M. Moles\inst{5,16}
        \and
        L. Sodr\'e Jr.\inst{19}
        \and
        K. Taylor\inst{20}
        \and
        J. Varela\inst{16}
        \and
        H. V\'azquez Rami\'o\inst{16}
        \and JPAS team
          }
   \institute{CAS Key Laboratory for Research in Galaxies and Cosmology, Shanghai Astronomical Observatory, CAS, Shanghai, 200030, China.
         \and
             Instituto de Física, Universidade Federal do Rio Grande do Sul (UFRGS), Av. Bento Gonçalves, 9500, Porto Alegre, RS, Brazil
        \and
            Department of Astronomy, Tsinghua University, Beijing 100084, Peopleʼs Republic of China     
        \and
            Donostia International Physics Center, Paseo Manuel de Lardizabal 4, E-20018 Donostia-San Sebastian, Spain
        \and
            Instituto de Astrof\'isica de Andaluc\'ia (CSIC), P.O. Box 3004, 18080 Granada, Spain
        \and
            Observat\'orio Nacional/MCTI, Rua General Jos\'e Cristino, 77, São Crist\'ov\~ao, CEP 20921-400, Rio de Janeiro, Brazil 
        \and
            Departamento de F\'isica Matem\'atica, Instituto de F\'{\i}sica, Universidade de S\~ao Paulo, Rua do Mat\~ao, 1371, CEP 05508-090, S\~ao Paulo, Brazil
        \and
            Ethiopian Space Science and Technology Institute (ESSTI), Entoto Observatory and Research Centre (EORC), Astronomy and
            Astrophysics Research and Development Department, Addis Ababa, Ethiopia.
        \and
            College of Astronomy and Space Sciences, University of the Chinese Academy of Sciences, Beijing 100049, China 
        \and
            Sydney Institute for Astronomy, School of Physics A28, The University of Sydney, Sydney, NSW 2006, Australia 
        \and
            Instituto de F́ısica de Cantabria, CSIC-Universidad de Cantabria, E-39005 Santander, Spain 
        \and
            Department of Theoretical Physics, University of the Basque Country UPV-EHU, 48040 Bilbao, Spain
        \and
            Donostia International Physics Center (DIPC), 20018 Donostia, The Basque Country and IKERBASQUE, Basque Foundation for Science, Alameda Urquijo, 36-5 48008 Bilbao, Spain
        \and
            Ikerbasque, Basque Foundation for Science, E-48013 Bilbao, Spain
        \and
            Instituto de F\'isica, Universidade Federal da Bahia, 40210-340, Salvador, BA, Brazil
        \and
            Centro de Estudios de Física del Cosmos de Aragón (CEFCA), Plaza San Juan, 1, E-44001, Teruel, Spain
        \and
            Department of Astronomy, University of Michigan, 311 West Hall, 1085 South University Ave., Ann Arbor, USA
        \and
            University of Alabama, Department of Physics and Astronomy, Gallalee Hall, Tuscaloosa, AL 35401, USA
        \and
             Universidade de S\~ao Paulo, Instituto de Astronomia, Geof\'isica e Ci\^encias Atmosf\'ericas, Depto. de Astronomia, Rua do Mat\~ao, 1226, CEP 05508-090, S\~ao Paulo, Brazil
        \and
            Instruments4, 4121 Pembury Place, La Canada Flintridge, CA 91011, USA
             }
\abstract
   {The Ly$\alpha$ emission is an important tracer of neutral gas in a circum-galactic medium (CGM) around high-z quasi-stellar objects (QSOs). The origin of Ly$\alpha$ emission around QSOs is still under debate, bringing on significant implications for galaxy formation and evolution.}
   { In this paper, we study Ly$\alpha$ nebulae around two high redshift QSOs, SDSS J141935.58+525710.7 at $z=3.218$ (hereafter QSO1) and SDSS J141813.40+525240.4 at $z=3.287$ (hereafter QSO2), from the miniJPAS survey within the All-wavelength Extended Groth Strip International Survey (AEGIS) field.}
   {Using the contiguous narrow-band (NB) images from the miniJPAS survey and Sloan Digital Sky Survey (SDSS) spectra, we analyzed their morphology, nature, and origin.}
   {We report the serendipitous detection of double-core Ly\al\ morphology around two QSOs, which is rarely seen among other QSOs.
   The separations of the two Ly\al~cores are 11.07 $\pm$ 2.26 kpcs (1.47 $\pm$ 0.3$^{\prime\prime}$) and 9.73 $\pm$ 1.55 kpcs (1.31 $\pm$ 0.21$^{\prime\prime}$), with Ly$\alpha$~line luminosities of $\sim$ 3.35 $\times 10^{44}$ erg s $^{-1} $ and $\sim$ 6.99 $\times$ 10$^{44}$ erg s $^{-1}$ for QSO1 and QSO2, respectively. The miniJPAS NB images show evidence of extended Ly\al~ and CIV morphology for both QSOs and extended HeII morphology for QSO1.}
   {These two QSOs may be potential candidates for the new enormous Lyman alpha nebula (ELAN) found from the miniJPAS survey due to their extended morphology in the shallow depth and relatively high Ly$\alpha$ luminosities. We suggest that galactic outflows are the major powering mechanism for the double-core Ly$\alpha$ morphology.
   Considering the relatively shallow exposures of miniJPAS, the objects found here could merely form the cusp of a promising number of such objects that will be uncovered in the upcoming full Javalambre-Physics of the Accelerated Universe Astrophysical Survey (J-PAS) survey and deep integral field units (IFU) observations with 8-10m telescopes will be essential for constraining the underlying physical mechanism that is responsible for the double-cored morphology.}
\maketitle

%
%
\section{Introduction}
The Lyman-alpha line of neutral hydrogen (Ly\al~$\lambda$1215.67) at high redshift is a powerful probe of the Epoch of Reionization \citep[EoR, e.g.,][]{2006Malhotra, 2006Fan, 2011Dijkstra, 2013Treu, 2015Mesinger, 2018Davies, 2018Mason, 2019Greig}. It is a gas tracer in the circum-galactic medium (CGM) and intergalactic medium (IGM) that can be used to characterize the IGM ionization state and galaxy properties. The CGM is the gaseous component that regulates the gas exchange between the galaxy and the IGM. Both the IGM and CGM are vast reservoirs responsible for fueling the star formation of galaxies. Therefore, studies of the CGM are crucial for understanding galaxy formation and evolution, especially in the early Universe \citep{Tumlinson, 2020Peroux}. Moreover, the Ly\al~line can be used to trace circum-galactic gas around high redshift star-forming galaxies \citep{2000Steidel, 2004Matsuda, 2016Wisotzki, 2018Wisotzki, 2020Herenz, 2022Kusakabe}, radio galaxies \citep{2006Miley, 2019Marques-Chaves, 2021Shukla}, and quasars \citep[QSOs,][]{Cantalupo, 2016Arrigoni, 2016Borisova, Cai2019, 2019Arrigoni, 2022Costa}  as well as Ly\al~halos or nebulae.

Several phenomena that have been proposed to explain the origins of Ly\al~emission, including photoionization by active galactic nuclei \citep[AGN,][]{2001Haiman, 2009Geach, Cai2017}, shock heating by galactic superwinds \citep{2000Taniguchi}, cooling radiation from cold-mode accretion \citep{2000Haiman, 2001Fardal, 2005Furlanetto, 2009Dijkstra, 2010Goerdt,2010Faucher-Gigu, 2012Rosdahl}, and resonant scattering of Ly\al~photons from star-forming galaxies \citep{2018Verhamme, 2019Claeyssens, 2020Leclercq, 2021Chen} or AGN in the embedded galaxies \citep{2011Hayes, 2011Steidel}. It has also been proposed that a combination of these mechanisms could also play a role \citep{2005Furlanetto, 2015Lake, Ouchi2020, 2021Byrohl, 2021Garel, 2021Mitchell}. However, the physical drivers responsible for the diffuse Ly\al~emission are still under debate.

Typical Ly\al~emission around QSOs is diffuse and characterized by very low surface brightness (SB) levels, e.g., 10$^{-18}$ - 10$^{-20}$ erg s$^{-1}$ cm$^{-2}$ arcsec$^{2}$ \citep{1996Haardt, 2005Cantalupo, 2008Rauch, 2016Borisova, 2016Arrigoni}.
Therefore, detecting the full extent of very faint Ly\al~emission is an extreme challenge and current facilities are not yet capable of detecting such low SB levels. Several observational efforts have been made to search for diffuse Ly\al~around high redshift QSOs. Previous detections have employed different techniques such as narrow-band (NB) imaging \citep[e.g.,][]{2004Francis, 1987Hu}, long-slit spectroscopy \citep[e.g.,][]{2003Bunker, 1991Heckman, 2012North}, or integral field spectroscopy \citep[IFS, ][]{2016Borisova, 2006Christensen} for the detection of Ly\al~emission around QSOs. Among those three techniques, IFS is more efficient in the wavelength coverage while often constrained in a much smaller field of view. Most of the large Ly\al~nebulae like enormous Lyman alpha nebulae \citep[ELANs,][]{Cantalupo,2015Hennawi,Cai2017,2018Arrigoni} are detected by either long exposure NB stacking analysis (e.g., 18 hrs at the 4m-Kitt Peak National Observatory (KPNO)/MOSAIC-1.1; \cite{Cai2017}, 10 hrs at the 10m-Keck/ Low Resolution Imaging Spectrograph (LRIS); \cite{Cantalupo} or using the multi-unit spectroscopic explorer (MUSE)-integral field units (IFU) at the 8m Very Large Telescope (VLT) \citep[e.g., 4.5 hrs][]{2018Arrigoni} to achieve a SB level of 10$^{-19}$ erg s$^{-1}$ cm$^{-2}$ arcsec$^{2}$ to trace the extended diffuse emission. 

The NB surveys are often taken in a wide field to efficiently detect the diffuse Ly\al~emission while are often limited by fixed redshift windows.
The Javalambre-Physics of the Accelerated Universe Astrophysical Survey \citep[J-PAS,][]{2014Benitez} is a multi-band survey with 54 NB (and 5 BB) filters.
Because of its large field of view, the sub-arcsec spatial resolution, and contiguous NB filters acting as low-spectral resolution IFU, it is an ideal survey for studying the diffuse Ly\al~emission around high-redshift QSOs. 

In this paper, we present the detection of double-cored Ly\al~morphology of two high-z QSOs in the miniJPAS survey \citep{2021Bonoli}. MiniJPAS is a 1 deg$^2$ NB photometric survey on the AEGIS field, a proof-of-concept survey of the larger J-PAS project, which will soon start mapping thousands of deg$^2$ of the northern sky in 59 bands \citep[54 NBs and 5 BBs,][]{2021Bonoli}. This paper is structured as follows. We introduce the sources and the JPAS and miniJPAS observations in Section 2. Our analysis and results are presented in Section 3. A discussion on the physical properties and various powering mechanisms that contribute to the  Ly\al~emission of the studied objects is given in Section 4. The main conclusion is presented in Section 5. Throughout the paper, we convert redshifts to physical distances (with a scale of 7.535 kpc/${^{\prime\prime}}$ at $z=3.218$ and 7.482kpc/${^{\prime\prime}}$ at $z=3.287$) assuming a $\Lambda$CDM cosmology with  $\Omega_\mathrm{M}$ = 0.3 and $\Omega_{\Lambda} = 0.7$ and $H_{0} = 70~\mathrm{km~s^{-1}~Mpc^{-1}}$. All magnitudes are given in the AB system. 
\vspace*{-4mm}
\section{Observation and data}
\subsection{J-PAS and mini-JPAS surveys}

The Javalambre-Physics of the Accelerated Universe Astrophysical Survey \citep[J-PAS, ][]{2014Benitez} is a wide-area photometric survey soon to be conducted at the Javalambre Astrophysical Observatory (OAJ), located at the Sierra de Javalambre (Teruel, Spain). It is poised to obtain a deep, sub-arcsec spectrophotometric map of the northern sky across 8000\,deg$^{2}$. The survey will use the 2.5 m JST/T250 telescope and obtain multi-band imaging in optical bands with an effective field of view of 4.2\,deg$^{2}$ and a plate scale of 0.225 arcsec pixel$^{-1}$. 
The J-PAS filters provide low-resolution spectroscopy (resolving power, R $\sim$ 60;  also referred as J-spectra in this paper) of a large sample of astronomical objects with 54 NB filters (3780 to 9100 \AA; $\Delta$\lam =~145  \AA ), one blue medium-band (MB) filter (3497 \AA; $\Delta$\lam =~509\AA), one red MB filter (9316 \AA; $\Delta$\lam =~635\AA), and three Sloan Digital Sky Survey (SDSS) Broad-Band (BB) filters. J-PAS nominal depth at signal-to-noise (S/N)$\sim$ 5 is between 22 to 23.5 AB mag for the NB filters and 24 for BB filters \cite{2021Bonoli}. With a large set of NB filters, J-PAS makes it suitable for an extensive search for extended Ly\al~emission around QSOs from redshifts $z=2.11$ to 6.66. Meanwhile, the wide spectral coverage of the J-spectra can cover Ly\al~, SIV+OIV, CIV, HeII, CIII], CII, and Mg II at $z>2$, which allows us to study the properties of high redshift QSOs. 
 
The miniJPAS survey \citep{2021Bonoli} observed four All-wavelength Extended Groth Strip International Survey \citep[AEGIS;][]{2007Davis} fields in 2018-2019 with the pathfinder camera and detected more than 60,000 objects in a total area of 1 deg$^{2}$. The miniJPAS survey demonstrates the capabilities and unique potential of the upcoming J-PAS survey \citep{2021Hern, 2021Gonz, 2021Mart, 2022Queiroz, 2022Mart, 2022Chaves-Montero}. More details about the observation and data reduction of miniJPAS are given in \cite{2021Bonoli}. We obtained the miniJ-PAS images, flux catalogs, filter curves, and all other instrument information of these QSOs from the J-PAS database (CEFCA archive)\footnote{http://archive.cefca.es/catalogues/minijpas-pdr201912}. 
\subsection{Source properties}

\begin{figure*}[htb!] 
   \centering
\includegraphics[width=0.9\textwidth]{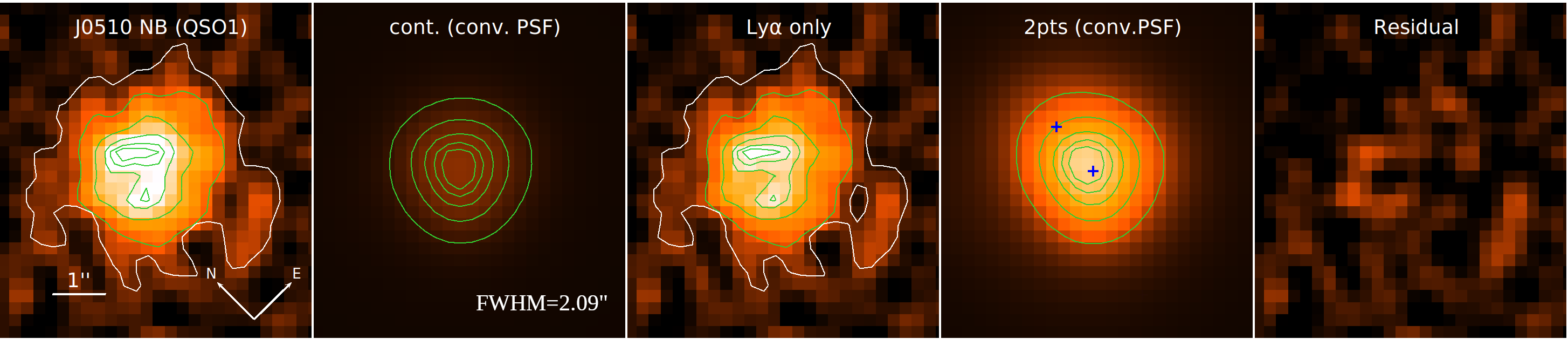}
\includegraphics[width=0.9\textwidth]{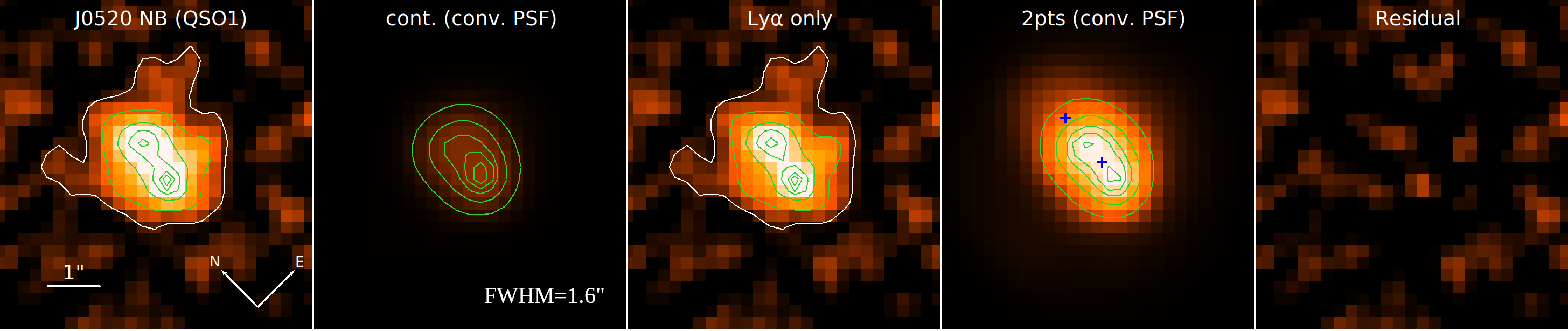}

\vspace{0.2cm}

\includegraphics[width=0.9\textwidth]{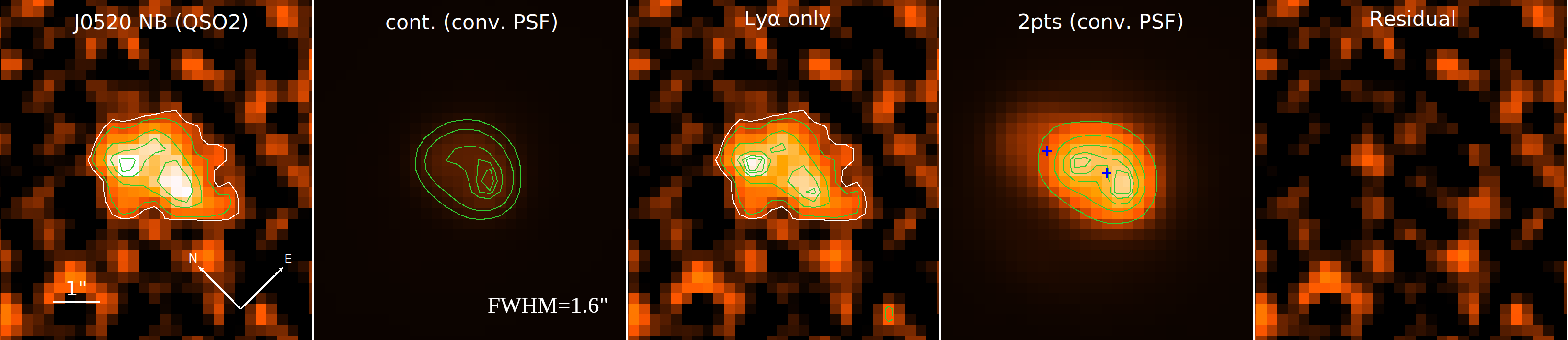}

\hspace{0.022cm}
\includegraphics[width=0.9\textwidth]{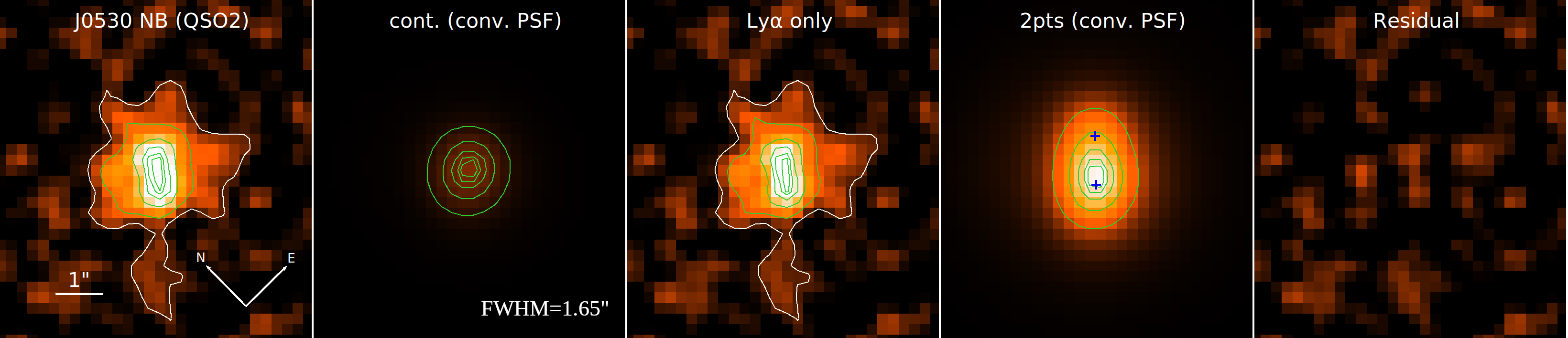}
   \caption{QSO1 and QSO2 in two Ly\al~NB J-PAS filters. First column shows NB miniJPAS image; second shows continuum image created by convolving PSF of Ly\al~image with continuum flux; third column is the continuum subtracted Ly\al~image; fourth column is the PSFs in two points convolved with Ly\al~flux; and fifth column shows the residual after subtracting two pts PSF from Ly\al~image. Blue crosses represent the center of the PSF position given for GALFIT fitting, and the white contour in the 1st and 3rd column represents an isophote of $2\sigma$ above the background STD. Green contours represent 30\%, 60\%, 80\%, 90\%, and 95\% of the peak values in the corresponding images. The FWHM size of the PSFs is indicated in the second column.}
  \label{fig:lya_im}
\end{figure*}
\begin{figure}[htb!] 
\includegraphics[width=0.48\textwidth]{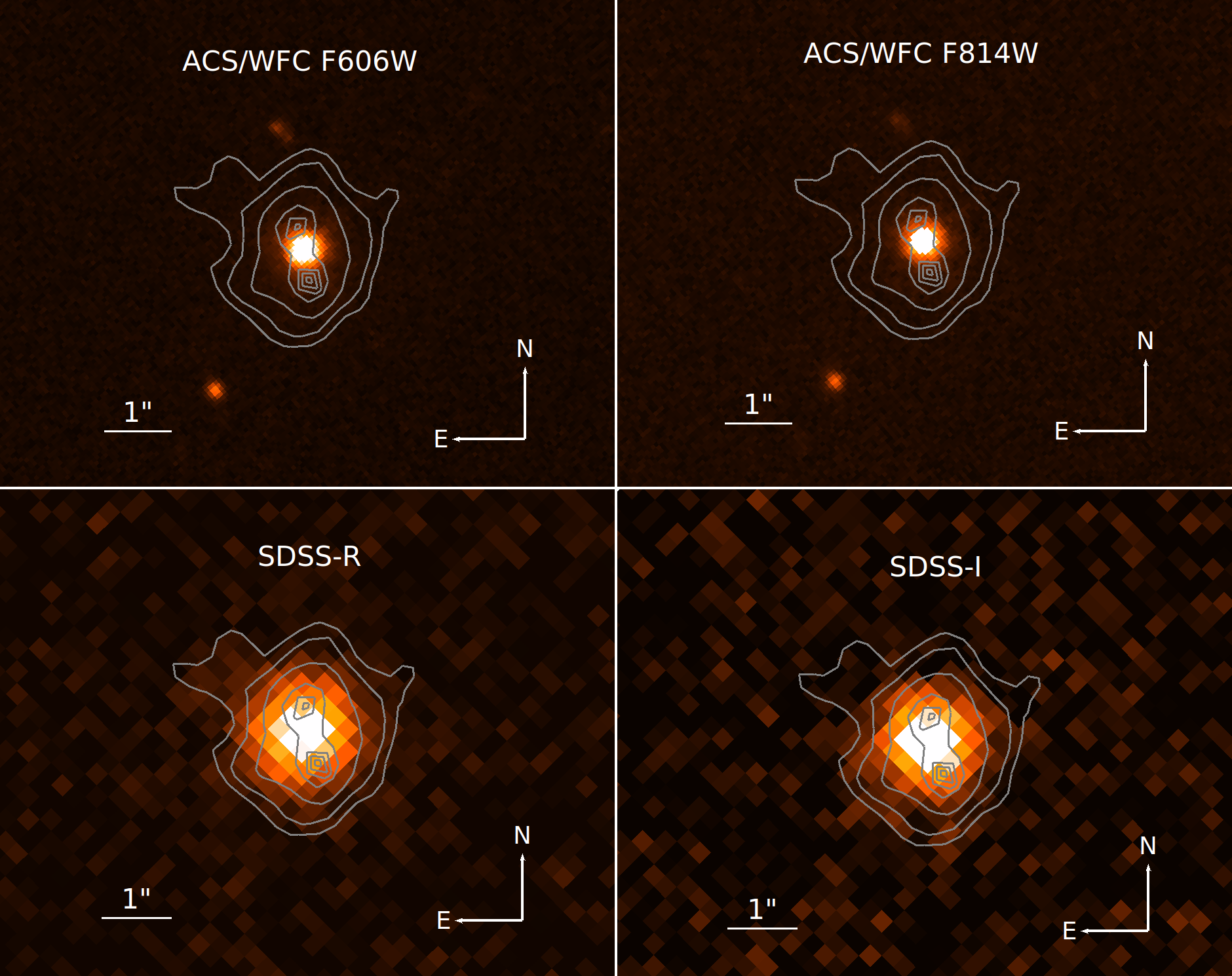}
\caption{HST images of QSO1 in HST/ACS F606W and F814W filters. The grey contours represent contour levels from JPAS/J0520 NB filter (2, 4, 8, 14, 16, 17, and 18 $\sigma$ levels of background STD).}
  \label{fig:hctim}
\end{figure}
With miniJPAS, we report two interesting QSOs, SDSS~J141935.58+525710.7 (RA: 214.8983, Dec: 52.953, hereafter QSO1) at a redshift of $z_\mathrm{spec}=3.218$ and SDSS~J141813.40+525240.0 (RA: 214.5559, Dec: 52.8778, hereafter QSO2) at redshift $z_\mathrm{spec}=3.287$, with the double-core Ly\al~morphology (Fig.~\ref{fig:lya_im}). We selected these two QSOs in a systematic search on the Ly\al~nebulae around 59 (SDSS and miniJPAS cross-matched targets) spectroscopically confirmed high redshift QSOs ($z>2$) using miniJPAS (Rahna et al., in prep).
In the first step, we selected spectroscopically confirmed QSOs at z$>$2 from SDSS survey DR16 and cross-matched them with miniJPAS 4 AEGIS fields, yielding 61 QSOs and after discarding the QSOs with bad warning flag of spectra, the final sample consists of 46 QSOs at 2$<$z$<$4.29. We utilized the contiguous NB imaging of miniJPAS for our selection criteria of Ly\al~nebulae. Our target selection of Ly\al~nebulae is based on the excess of Ly\al~flux and morphology in the filter covering the Ly\al~line and continuum. A QSO with a size of Ly\al~morphology greater than its continuum morphology at a 2 sigma level of Background is considered a Ly\al~nebula, after accounting for its point spread function (PSF). PSFs used in the analysis are downloaded from the JPAS database, where PSFs are created from PSFex (Moffat profiles) using nearby stars. Only these two targets from the parent sample show double-core morphology in their Ly\al~ NBs. There are four other QSOs in the same redshift range of these two targets and their Ly\al~morphology doesn't show double-core morphology.
More details about the parent sample are explained in Rahna et al., in prep. Spectroscopic redshifts were available for these two QSOs thanks to SDSS/BOSS \citep{2013Dawson,2017Paris}.

QSO1 has broadband observations with the Hubble Space Telescope (HST) Advanced Camera for Surveys (ACS/WFC; AEGIS Survey \citep{2007Davis}). The HST/ACS F606W and F814W wide band images of QSO1 show a single point-like morphology (Fig.~\ref{fig:hctim}). 
The X-ray image of QSO1 from stacked Chandra observation of 800 ks (AEGIS-X survey \citep{2009Laird, 2015Nandra} and the XDEEP2 survey \citep{2012Goulding}) shows QSO1 as a point-like source \footnote{QSO1 was located at $\approx$7 arcmin from the aimpoint of ACIS-I: at that distance, the ACIS-I PSF is elongated and distorted, with a 50\% radius of about 2 arcsec (= 50\% of the counts fall within 2 arcsec). Combining this large PSF with the moderately small number of counts (about 500 counts in the stacked image), means that some faint, extended X-ray structures on scales $<$2 arcsec would be easily missed, especially with automated classifications of sources in large surveys.}.
The radio observation of QSO1 with VLA (Very Large Array) in 6cm band \citep{2007Ivison} shows weak but single-source detection. QSO2 has not been observed by HST, Chandra, XMM-Newton (X-ray Multi-mirror Mission), nor VLA.

\section{Results}
Here, we discuss the analysis and results of the detection of the double-core Ly\al~morphology of the two high-redshift QSOs. In Section 3.1, we introduce the detection of the double-core Ly\al~nebulae in the miniJPAS survey. We then compare the J-spectra with the available SDSS spectra of the two QSOs in Section 3.2. We calculate the emission line luminosities and line ratios of the two QSOs in Section 3.3.

\begin{table*}
\begin{center}
\caption{Parameters for GALFIT PSF fitting\label{Tab:cores}}
\resizebox{\textwidth}{!}{
\begin{tabular}{cccccccccc}
\hline
Source & Core & \multicolumn{2}{c}{Position} & Separation & Magnitude & $\chi^{2}$ & 2$\sigma$ & Peak & SB at peak \\
 &  &  RA (ICRS)&  Dec(ICRS) &
($\prime\prime$/kpc)  &(mag)&  & (ADU) & ($\sigma$ level)& ( erg s$^{-1}$ cm$^{-2}$ arcsec$^{2}$)\\
\hline
{QSO1-J0510} & {North Core}& {14:19:35.631}&{+52:57:12.019} & {1.56 $\pm$ 0.31/11.75 $\pm$ 2.34} & {21.65 $\pm$ 0.09} & {0.792} & {0.0178} & {26} & {1.76 $\times$ $10^{-15}$} \\
{} & {South Core}& {14:19:35.597}&{+52:57:10.790} & {} &  {19.75 $\pm$ 0.01} &  {} & {} &{25} & {1.69 $\times$$ 10^{-15}$} \\
{QSO1-J0520} & {North Core}& {14:19:35.631}&{+52:57:12.064} & {1.47 $\pm$ 0.3/11.07 $\pm$ 2.26} & {21.71 $\pm$ 0.12} & {1.123} & {0.0284} &{17} & {2.97 $\times$$ 10^{-15}$} \\
{} & {South Core}& {14:19:35.585}&{+52:57:10.586} & {} & {19.83 $\pm$ 0.02} & {} & {} &{18} & {3.29 $\times$$ 10^{-15}$}  \\
{QSO2-J0520} & {North Core}& {14:18:13.385}&{+52:52:41.162} & {1.312 $\pm$ 0.207/9.73 $\pm$ 1.55} & {20.42 $\pm$ 0.03} & {1.067} & {0.0271}  & {9} & {1.48 $\times$$10^{-15}$} \\
{} & {South Core} & {14:18:13.442} & {+52:52:39.952} & {} & {22.00 $\pm$ 0.17}  & {} &{} & {9} & {1.48 $\times$$ 10^{-15}$} \\
{QSO2-J0530} & {North Core} & {14:18:13.486} & {+52:52:40.535} & {0.986 $\pm$ 0.21/7.38 $\pm$ 1.57} & {21.75 $\pm$ 0.06} & {1.097} & {0.0124} & {21} & {1.1 $\times$$ 10^{-15}$} \\
{} & {South Core} & {14:18:13.413} & {+52:52:39.801} & {} & {21.10 $\pm$ 0.02}  & {} & {} & {21} & {1.1 $\times$ 10$^{-15}$} \\
\hline
\end{tabular}
}
\small
Column (1): Filter name of the QSOs, Column (2-7): Cores, input PSF position in GALFIT, the separation between two PSFs, best fit magnitude and $\chi^{2}$ values from GALFIT fitting. Column (8-10): the value of 2 $\sigma$ (STD of background), the significance of the detection in $\sigma$ level compare to the background level, and corresponding SB values.
\end{center}
\end{table*}
\begin{table*}
\begin{center}
\caption{Properties of the emission lines detected in the J-spectra of QSO1 and QSO2\label{Tab:NBfilt}}
\resizebox{\textwidth}{!}{
\begin{tabular}{c|c|c|c|c|c|c|c|c|c|c|c|c}
  \hline
Source & Em.line & $\lambda_{Res}$ &  J-PAS & $\lambda_{eff}$ & $\Delta\lambda$ & Depth  &
\multicolumn{2}{c}{APER size} & 
$L_{\lambda,APER2}$ (miniJPAS) &
$L_{\lambda,2\sigma}$(miniJPAS) &
 $L_{\lambda,filtconv}$(SDSS) & $L_{\lambda,linespec}$(SDSS)  
  \\
 & & ($\AA$) &  Filter & ($\AA$) & ($\AA$) & (mag) &($^{\prime\prime}$) & (kpc) &(10$^{43}$ erg s$^{-1}$ )& (10$^{43}$ erg s $^{-1} $ )& (10$^{43}$ erg s $^{-1} $ ) & (10$^{43}$ erg s $^{-1} $ ) \\
  \hline
QSO1 & Ly\al~& 1215.67 & {J0510} & {5097} & {148} & 24.37 & {6} & {45.21}   & {20.73$\pm$2.44} & {69.97$\pm$6.767}  &{37.15$\pm$0.39} &  {62.41$\pm$0.64} \\  
&{Ly\al~+NV}& {1215.67} & {J0520} & {5202} & {150} & 23.48 & {4} & {30.14}  &      {32.92$\pm$2.85} & {48.64$\pm$3.84} & {24.93$\pm$0.38} &{19.34$\pm$0.72} \\
 & {SIV+OIV} & {1399.8} &  {J0590} &  {5917} &  {152} & 23.39 & {3} &  {22.61} & {2.57$\pm$1.74} &   {8.59$\pm$2.54}   &{3.84$\pm$0.31} & ... \\
 &{CIV} & {1549.48} &  {J0650} &  {6506} &  {146} & 24.06 & {6} & {45.21}  &     {12.15$\pm$0.69} &{19.227$\pm$2.39} &{12.48$\pm$0.27} & {16.4$\pm$0.44} \\
 &{HeII} & {1640.42} &  {J0690} &  {6912} &  {148} & 23.75 & {4} &  {30.14} &   {1.26$\pm$0.98}& {2.45$\pm$2.05}   & {1.35$\pm$0.25} &{1.77$\pm$0.32} \\
 &{CIII]} & {1908.734} &  {J0800} &  {8009} &  {140} & 22.89 & {3} &  {22.61}   &   {4.55$\pm$1.47}   &{5.54$\pm$2.29}  & {4.07$\pm$0.198}&  {6.76$\pm$0.37} \\ \cline{1-13}
QSO2 & Ly\al~ & {1215.67} & {J0520} & {5202} & {150} & 23.48& {3} & {22.45}   & {22.52$\pm$2.31}& {33.53$\pm$3.46}  & {23.05$\pm$0.38} &{28.13$\pm$0.57} \\  
 &{Ly\al~+NV}& {1215.67} & {J0530} & {5296} & {150} & 24.59& {6} & {44.89}  &     {12.85$\pm$ 1.6} &{21.79$\pm$3.49} & {11.29$\pm$0.37}& {12.27$\pm$0.82} \\
 &{SIV+OIV} & {1399.8} &  {J0600} &  {6010} &  {150} & 23.97 & {3} &  {22.45} &   {2.22$\pm$1.34} & {3.07$\pm$2.68} & {1.40$\pm$0.31}  &{...}  \\
 &{CIV} & {1549.48} &  {J0660} &  {6607} &  {151} & 24.97 & {4} &  {29.93} &     {7.32$\pm$0.76} &{10.45$\pm$1.52} & {10.08$\pm$0.26}&  {13.52$\pm$0.63} \\
 &{HeII} & {1640.42} &  {J0700} &  {7007} &  {148} &23.85  & {2} &  {14.96} &     {2.42$\pm$1.11} &{2.42$\pm$1.11} & {2.08$\pm$0.24} & {1.35$\pm$0.63} \\
 & {CIII]} & {1908.734} &  {J0820} &  {8226} &  {143} & 23.19 & {2} &  {14.96} & {2.18$\pm$1.6} & {2.18$\pm$1.6}  & {0.761$\pm$0.19} & {3.28$\pm$0.43} \\
 \hline
\end{tabular}
}
\small
Column (1): Name of the emission line, Column (2): rest-frame wavelength of the emission line, Column (3-7): properties of J-PAS filters-filter name, effective wavelength, bandwidth, magnitude at SNR$=$5 in 1 square arcsec, Column (8-9): Aperture size of J-PAS catalog flux based on the size of 2$\sigma$ isophote above background STD, Column (10-13): Line luminosity ($L_{\lambda}$) measured in APER2, 2$\sigma$ isophote size from miniJPAS images and J-PAS filter curves convolved with SDSS spectra and total line luminosity given in SDSS BOSS spectra.
\end{center}
\end{table*}
\subsection{Detection of spatially resolved double-core morphology in Ly\al~NB images}
The NB images covering the Ly\al~line for the QSOs analyzed in this paper show a serendipitous double-cored morphology of Ly\al~(Fig.~\ref{fig:lya_im}). The redshifted Ly\al~emission lines of the two QSOs fall into two NBs of J-PAS (J0510 and J0520 for QSO1 and J0520 and J0530 for QSO2). QSO1 shows features of double core morphology in both Ly\al~images, albeit more significant in J0520. QSO2 shows double-core morphology only in J0520. This may be due to the fact that J0530 covers mostly the NV line of QSO2 and the NV contamination may cause different morphology (Fig.~\ref{fig:spec}). Furthermore, J0530 also shows an elongated core at the center with extended morphology compared to its compact and circular PSF. On the other hand, the continuum image only shows a compact single-core morphology. We find that the double-core morphology is not caused by the PSF although it is elongated in the J0520 filter. This also confirms with the morphology of nearby stars, which only exhibit single-core morphology with elongated PSFs. Based on the morphology in the continuum bands, especially the single point structures revealed in HST F606W and F814W images (Fig.~\ref{fig:hctim}), we infer that the double-core structures in the J0520 image of the two QSOs are caused by the diffuse and resonant-scattered Ly\al~radiation (see Section 4). 

To quantitatively analyze the double-core morphology of the Ly\al~image of the two QSOs, we modeled the double-core with two PSF profiles through GALFIT \citep{2002Peng}. First, we subtracted the continuum flux from the NB image by using a PSF (of the NB image) scaled to the continuum flux to create the Ly\al~only image. The continuum value is taken from the nearby filter with the best SNR, which is consistent with the convolution of the SDSS spectra and the JPAS filters. Then, we convolved the two PSFs with the Ly\al~flux and subtract it from the Ly\al-only image to get the residual (Fig.~\ref{fig:lya_im}). The contour levels of the model created with two PSFs clearly show the double-core morphology compared to the actual elongated single-core PSF. This demonstrates that the double-core is not due to the PSF effects. Input and output parameters of GALFIT fitting are given in Table~\ref{Tab:cores}. We note that the center of the two GALFIT input PSFs (blue crosses) in J0520 doesn't coincide with the peak of the Ly\al~emission in the NB images because of the distorted structure of the PSF. Two cores are separated by a distance of 11.07$\pm$2.26 kpc in QSO1 and 9.73$\pm$1.55 kpc by QSO2 and SB peak at 18 and 9 sigma level of background standard deviation (STD).

Besides the double-core morphology in the Ly\al~ images, both QSOs also show diffuse and spatially extended Ly\al~radiation. The spatial extent was measured at $2\sigma$ isophote above the background STD (10$^{-16}$ -- 10$^{-17}$ erg s$^{-1}$ cm$^{-2}$ arcsec$^{2}$) from each image (Fig.~\ref{fig:jpasimages}). Compared to the size of the QSOs in the J0520 filter, QSO1 is more extended in J0510 ($\sim$ 45.21 kpcs), with a size difference of $\sim$ 15.07 kpcs, and QSO2 is more extended in J0530 ($\sim$ 44.89 kpcs), with a size difference of $\sim$ 22.44 kpcs (see APER size column of Table~\ref{Tab:NBfilt}). This may be due to the difference in the depth of the two filters (in Table~\ref{Tab:NBfilt}). However, the current miniJPAS has limited ability to estimate the actual size of Ly\al~nebulae due to shallow observation depth (480 secs) and PSF effects.
\vspace*{-1.5mm}
\subsection{J-spectra vs SDSS spectra}
\begin{figure*}
   \centering
\includegraphics[width=0.8\textwidth]{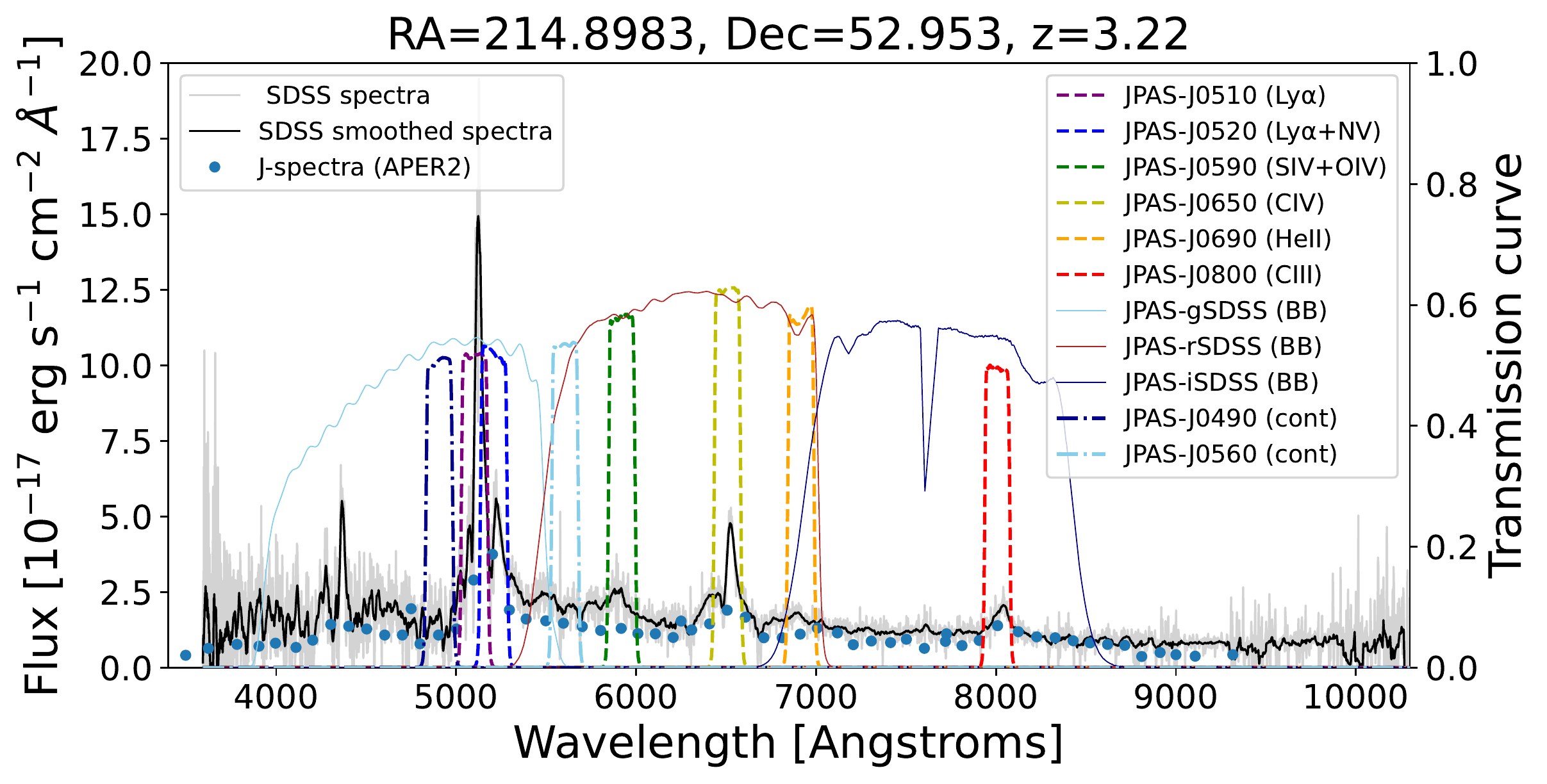}
\includegraphics[width=0.8\textwidth]{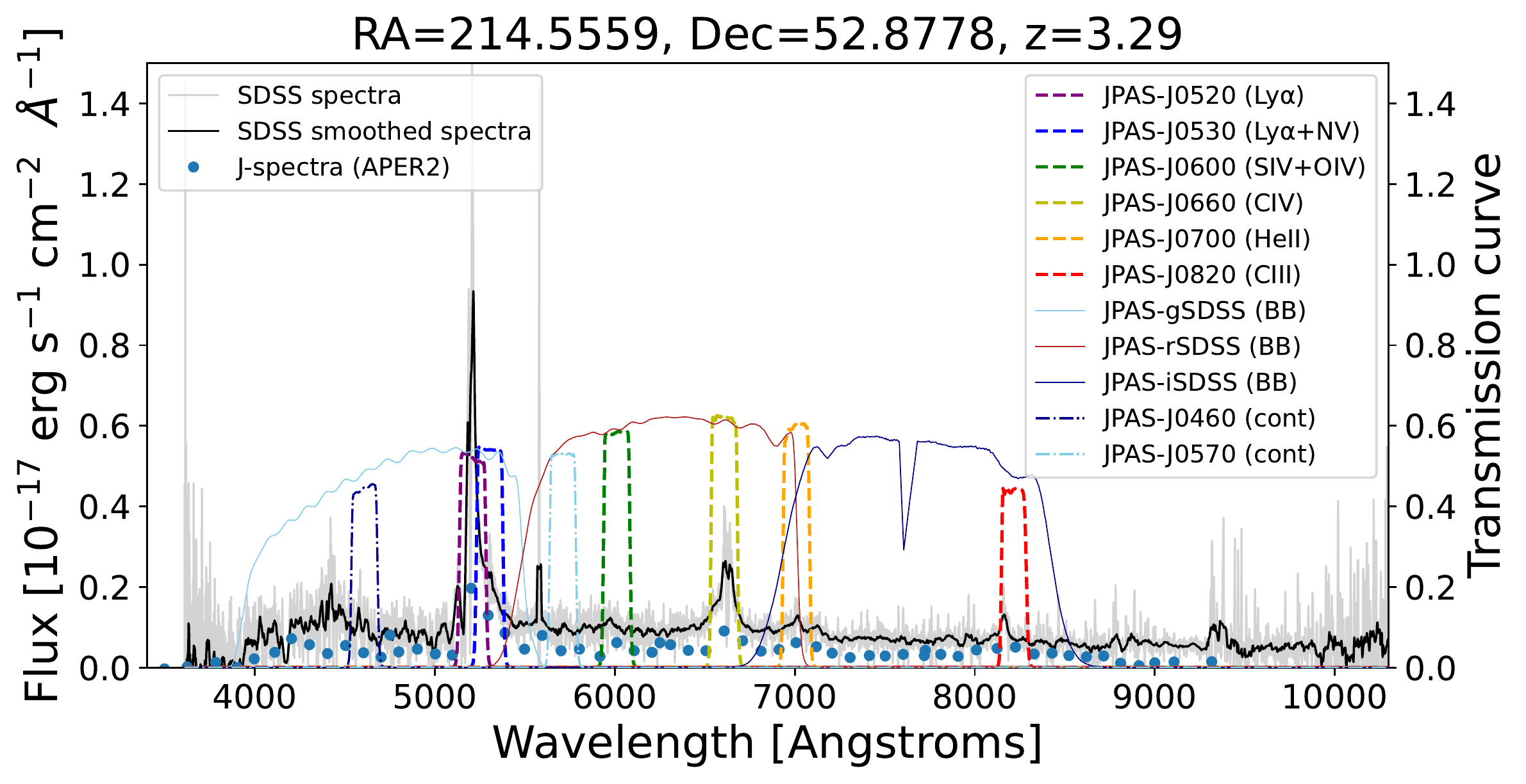}
\caption{SDSS spectra (grey), Smoothed SDSS spectra (black) and J-spectra (blue dots) overplotted with different NB and BB J-PAS filter curves covering the most prominent emission lines and the continuum (used to measure continuum flux of Ly\al~covering NB images).}
  \label{fig:spec}
\end{figure*}

We present the J-spectra (pseudo spectra) and the SDSS spectra of our two QSOs in the upper panels of Fig.~\ref{fig:spec}. The SDSS spectra are obtained from the final data release of Baryon Oscillations Spectroscopic Survey \citep[BOSS;][]{2013Dawson} in SDSS DR12 \citep{2017Paris}. QSO1 had been observed many times as part of the SDSS reverberation mapping project \citep{2019Hemler} (59 epochs from 2014 to 2020) and the spectrum in Fig.~\ref{fig:spec} was taken on March 19, 2019, while the  QSO2 spectrum was observed on June 6, 2013. In BOSS, the spectrum was observed through 2 arcseconds (diameter) fibers. The J-spectra in Fig.~\ref{fig:spec} is composed of the atmosphere-extinction corrected photometry (APER2-2$^{\prime\prime}$ diameter) of all J-PAS NB filters obtained from the CEFCA archive \citep{2019Lopez}. 
The prominent emission lines such as Ly\al, SIV+OIV, CIV, HeII, and CIII] detected in SDSS are visible in J-spectra. The properties of these filters are presented in Table~\ref{Tab:NBfilt}. These two QSOs are type 1 QSOs characterized by their broad emission lines. The spectroscopic Ly\al~profile of each QSO shows an asymmetric double-peaked profile and is dominated by the red peak, which is consistent with the general picture of IGM and CGM absorption \cite{2020Gurung}. The broadened Ly\al~profile indicates radiative transfer effects. The double peak in the CIV line of QSO1 may be due to the presence of CIV broad absorption lines (BALs), with high outflow velocity \citep[vmax
= 3243 km/s:][]{2019Hemler}. 

\begin{figure*}
   \centering
\includegraphics[width=1\textwidth]{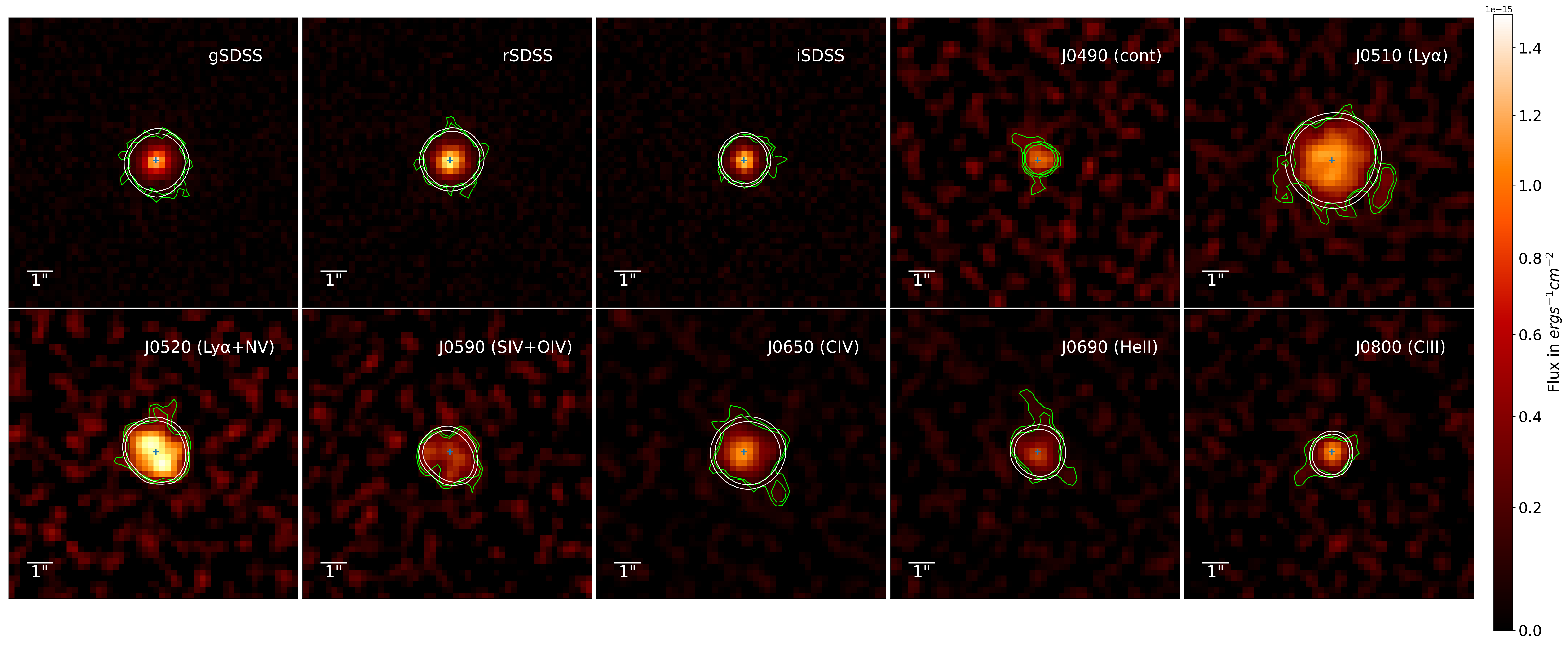}  
\includegraphics[width=1\textwidth]{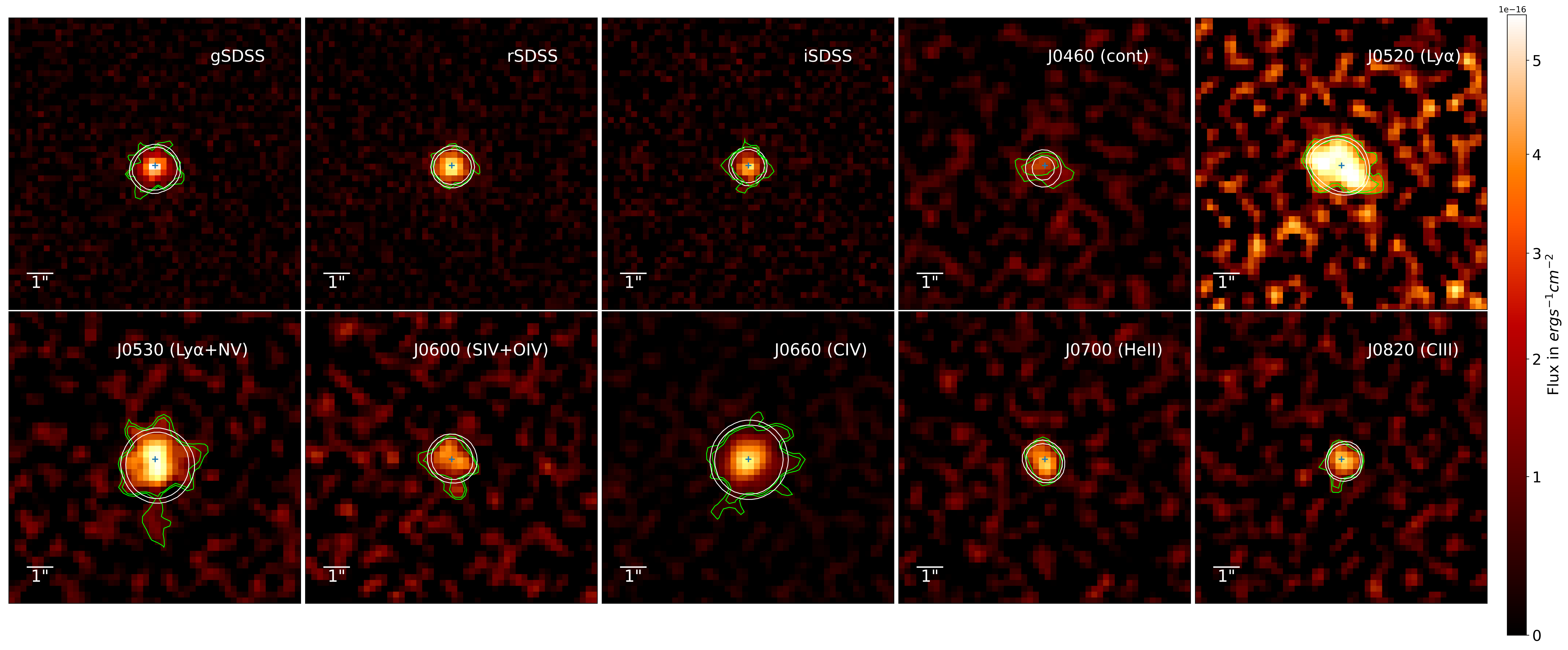}
 \caption{MiniJPAS BB images and NB images of QSO1 (upper panel) and QSO2 (lower panel) covering the most prominent emission lines. The cross is the center of the QSO. All images are $11.3^{\prime\prime} \times 11.3^{\prime\prime}$ size. The green (NB image) and white (PSF) contour levels represent the isophote of $2\sigma$ and $3\sigma$ above the background STD of the NB image. The center of the double-core coincides with the center of QSOs in BB filters.}
  \label{fig:jpasimages}
\end{figure*}

By utilizing miniJPAS contiguous NB imaging, we can study the emission line morphology in various wavelengths (as shown in Fig.~\ref{fig:jpasimages}). We note that other emission lines observed through the NB and BB images do not exhibit such prominent double-core structures as in the Ly\al~image.

 In Table~\ref{Tab:NBfilt}, we list the emission line filter properties, APER size used for estimating luminosity in Col. 11 and we estimate the emission-line luminosities from miniJPAS images and SDSS spectra. The APER size is the spatial extent measured at $2\sigma$ isophote above the background STD from each image. In addition to Ly\al, QSO1 also exhibits spatially extended emission in HeII (J0690), CIV (J0650), and CIII (J0690), while QSO2 shows extended morphology in CIV (J0660) (Fig.~\ref{fig:jpasimages}). The UV-continuum emission line (e.g., J0500, J0740) and BBs (gSDSS, rSDSS, iSDSS) are $\sim$ 15kpc in size in both QSO1 and QSO2 and do not show spatially extended emission, as in Ly\al~and CIV. The extended PSF in some emission line filters also contributes to the extended morphology. The comparison of 2$\sigma$ contours on NB image (green) with PSF (white) in Fig.~\ref{fig:jpasimages} shows how much the nebula is extended in different emission lines.
\subsection{Emission line luminosity and line ratios}
To derive the emission line luminosities of the extended nebulae from miniJPAS images, we adopted a method that exploits contiguous narrow band filters of J-PAS. In this method, we use two adjacent NB filters covering the emission line and the adjacent continuum (by assuming that the source continuum is approximately constant in the nearest NB filter covering the continuum) to calculate continuum-subtracted emission line fluxes. For Ly\al~, we used two filters (one on each side of the Ly\al~line: J0490, J0560 for QSO1 and J0460, J0570 for QSO2; Fig.~\ref{fig:spec}) to estimate the continuum because, at z$\sim$3, the blue part of the spectrum may be attenuated by the IGM \citep{2014Inoue} or any intervening NHI gas. The emission line luminosity estimated from such flux is tabulated in Table~\ref{Tab:NBfilt}. The spectral line widths of QSO1 and QSO2 are significantly wider than the filter width of J-PAS/miniJPAS NBs, which will result in a loss of flux in the luminosity calculated from miniJPAS images in a single filter comparison to SDSS spectra. The line luminosity calculated from SDSS spectra (by convolving J-PAS filter curves with SDSS spectra) and from miniJPAS NB images in the same aperture size of 2 arcsecs are within 1$\sigma$ error limit (of L calculated from miniJPAS), except L$_{Ly\alpha~}$and L$_{Ly\alpha~+NV}$ in QSO1 and L$_{CIV}$ in QSO2. This variation might be due to the PSF effects. The luminosity estimated within the 2$\sigma$ isophote size (col. 10) shows a significant increase in Ly\al~, SIV+OIV, CIV, and HeII (weak) in QSO1 and Ly\al~in QSO2 as aperture size increases. Besides any possible intrinsic variability, the different line flux measurements may also contribute to the flux difference.

The relative strength of the three important brightest UV emission lines such as Ly\al, CIV, and HeII provide a powerful probe of the thermodynamic properties of the gas, its ionization state, gas density, and metallicity. These emission lines can also disentangle the powering mechanism for the Ly\al~emission. For example, line ratio diagnostic diagrams can be used to distinguish the photo-ionization scenario and shock scenario \citep[e.g.,][]{1998Allen, 2015Arrigoni}. In the cooling zone behind high-velocity shocks, UV collisionally excited lines are strongly excited at high temperatures (2 $\times 10^{4}-10^{5}$ K), compared with photoionized plasmas, where these species are excited at $10^{4}$ K \citep{1998Allen}. Therefore, UV line intensities are predicted to be much stronger in shock scenarios than in simple photo-ionization scenarios. \cite{2015Arrigoni} demonstrated the CIV/Ly\al~versus HeII/Ly\al~diagram as a method to discriminate between shock and photoionization models with strong upper limits calculated above 10$^{-18}$ erg s$^{-1}$ cm$^{-2}$ arcsec$^{2}$ (at 2$\sigma$).

The line ratios of CIV/Ly\al~and HeII/Ly\al~are 0.262 $\pm$ 0.008, 0.028 $\pm$ 0.005 for QSO1 and 0.48 $\pm$ 0.024, 0.048 $\pm$ 0.022 for QSO2, which is estimated by the ratio of fluxes calculated from SDSS spectra. These values are consistent with the AGN photoionization (optically thick) and shock model for both QSOs explained by \cite{2015Arrigoni}. But we would need to reach SB level of 10$^{-18}$ to 10$^{-19}$ erg s$^{-1}$ cm$^{-2}$ arcsec$^{2}$ at 2$\sigma$ to trace the full extent of Ly\al~nebulae to discriminate these models. However, due to the limited depth (10$^{-16}$ erg s$^{-1}$ cm$^{-2}$ arcsec$^{2}$ at 2$\sigma$) and spatial-spectral power of miniJPAS, we cannot constrain further details about different models. A deep J-PAS like NB surveys or IFU observation with an 8-10m telescope would help to give a more definitive conclusion on this aspect.
 
\section{Discussions}
\begin{figure}[htb!]
\includegraphics[width=0.53\textwidth]{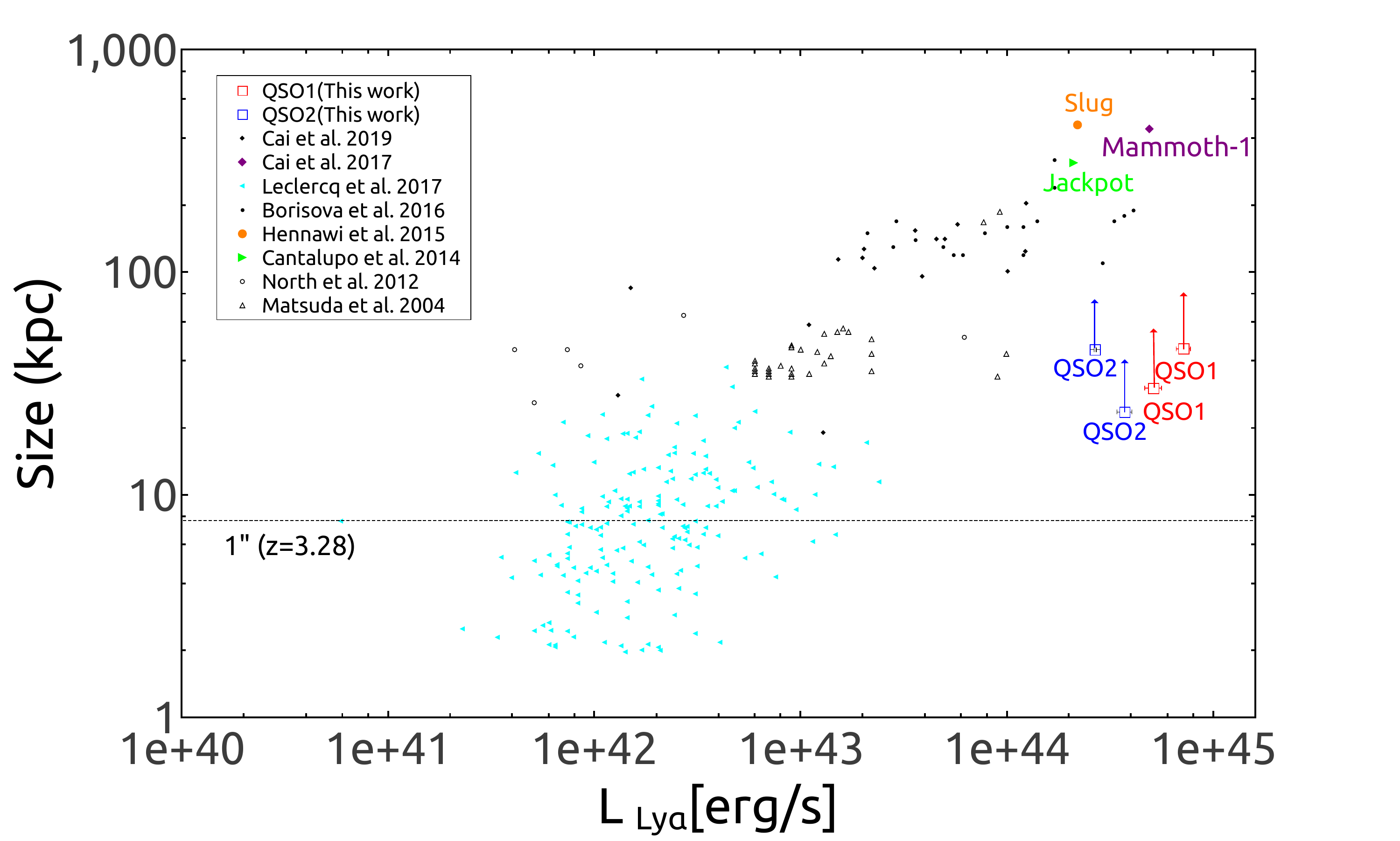}
   \caption{Comparison of Ly\al~luminosity and size of the Ly\al~nebulae detected in this study with Ly\al~nebulae from the literature. Two square symbols of QSO 1 and 2 indicate two Ly\al~ filters. The LAHs are in cyan color, LABs are in black color and other color symbols are ELANs. All these Ly\al~nebulae are at high redshift ($z>2$).
   We note that because of the different definitions of sizes, instrument sensitivity, methods, and redshifts, a direct comparison between the size and luminosity of the QSOs in this study with the literature is not possible.}
  \label{fig5:lumsize}
\end{figure} 
In this section, we discuss the implications based on the results presented in the previous section. In Section 4.1, we discuss the size and luminosity relations of the Ly\al~nebulae around the two QSOs and compare it with the literature candidates. We then discuss the physics behind the origin of the diffuse Ly\al~nebulae in Section 4.2 and the origin of the double-core Ly\al~ morphology in Section 4.3

\subsection{L$_{Ly\alpha}$ versus Size$_{Ly\alpha}$: Suggesting two possible new ELANs}
There is no clear definition for different types of Ly\al~nebulae. It can be roughly classified into Ly\al~halos (LAH), Ly\al~blobs (LABs), and enormous Ly\al~nebulae (ELAN) based on their luminosity and size in the Ly\al~wavelength \citep{Ouchi2020}. The LAHs are spatially extended Ly\al~nebulae with a physical scale of 1 to 10 kpc and a Ly\al~luminosity of $\sim 10^{42}-10^{43}$ erg s $^{-1} $ \citep{2004Hayashino, 2014Momose, 2016Momose, 2020Guo}, LABs have a physical scale of $\sim$ 10 -100 kpc and a luminosity of  10$^{43}-10^{44}$ erg s $^{-1}$ \citep{2000Steidel,2004Matsuda, 2018Shibuya, 2020Herenz}, and ELANs are the most luminous ($> 10^{44}$ erg s $^{-1}$) and extend to a large-scale of several hundreds of kpc (larger than the virial radius of dark matter halo) around $z>2$ QSOs \citep{Cantalupo, 2015Hennawi, Cai2017, 2018Arrigoni, 2022Nowotka}. 
 Based on these classifications Ly\al~nebulae around two QSOs come under bright LABs with a Ly\al~luminosity of $\sim 3-6 \times 10^{44}$ erg s $^{-1} $ and physical size of $\sim$ 30 -- 45 kpc (Fig.~\ref{fig5:lumsize}). These QSOs are the most luminous and have a similar range of luminosity as ELANs (e.g., Slug, Jackot, MAMMOTH-1 in Fig.~\ref{fig5:lumsize}). Most importantly the spatial extent of the Ly\al~nebulae strongly depends on the limiting surface brightness. \cite{2021Kimock} demonstrated Ly\al~luminosity and enclosed area as a function of limiting surface brightness through high-resolution cosmological zoom-in simulations. Comparing with their simulation results suggest that the current miniJPAS depth can only trace a small area near the center of nebulae and these two nebulae maybe two new ELANs.  
 Much deeper NB imaging is needed to confirm whether these objects are new ELANs.

\subsection{Physics of the extended Ly\al~morphology}
Here, we discuss several mechanisms, including resonant scattering \citep[e.g.,][]{2011Hayes, 2011Steidel}, gas photo-ionization by central AGN \citep[e.g.,][]{2001Haiman, 2009Geach, 2011Hayes}, shocks due to gas flows \citep[e.g.,][]{2000Taniguchi}, and cold gas accretion \citep[e.g.,][]{2000Haiman, 2001Fardal, 2005Furlanetto, 2009Dijkstra, 2012Rosdahl}, which could be responsible for the extended Ly\al~emission around QSOs. A combination of all these mechanisms may also act together to power Ly\al~nebulae.

The morphology of rest-frame UV lines such as CIV \lam 1549 and HeII \lam 1640 along with Ly\al~ provide additional information about different physical mechanisms explained in the previous section. For example, CIV/Ly\al~ratio provides a diagnostic for the halo gas metallicity and ionization state of halo gas in the CGM, while extended CIV morphology indicates the size of the metal-enriched halos from galactic outflows from central AGN \citep{2015Arrigoni}. Photoionization and galactic outflows are the two possible reasons for the Ly\al~emission if there is a detection of extended emission from both CIV and HeII lines. However, the UV line ratios (e.g., CIV/Ly\al~and HeII/Ly\al~) are different in both scenarios.  
If the Ly\al~nebulae is powered by shocks due to galactic-scale outflows or a shell-like or filamentary morphology with large Ly\al~ width of $\sim$ 1000 km/s and enormous Ly\al~($\sim$ 100 kpc) size is expected \citep{2000Taniguchi, 2001Taniguchi, 2003Ohyama, 2005Wilman, 2006Mori}. Both QSOs have a large Ly\al~ width of $>$ 1000 km/s, suggesting that shell or filamentary-like morphology could be expected for the extended Ly\al~emission surrounding them. However, the nominal depth of miniJPAS data cannot trace the extended size of these nebulae. Additionally, the NB covering the CIV emission line in both QSOs show extended morphology which suggests that the neutral gas expelled by outflows that are carbon (metal) enriched.

In gravitation cooling radiation, collisionally excited CGM gas emits Ly\al~photons and converts gravitational energy into thermal and kinetic energy as it falls into the dark matter (DM) halo potential. The luminosity of Ly\al~nebulae produced by cooling radiation (in a 10$^{11}$ DM halo) predicted to be $\leq$ 5$\times 10^{41}$ erg s $^{-1} $ through the hydrodynamic simulation of \cite{2009Dijkstra} and \cite{2012Rosdahl}. If Ly\al~nebulae has a luminosity higher than this value, it suggests that the mechanism producing Ly\al~emission is not only due to gravitation cooling radiation. Additionally, the powering mechanism for Ly\al~emission due to gravitational cooling radiation is not expected to produce extended CIV emission, but HeII emission can be expected \citep{2015Arrigoni}. This implies that the increased Ly\al~luminosity and extended CIV emission in both QSOs exclude a significant contribution from gravitational cooling.

Resonant scattering does not contribute to the large scale SB of  ($>$ 100kpc) of Ly\al~nebulae because resonantly scattered photons can escape the system effectively only in very small scales ($<$ 10kpc) \citep{Cantalupo, 2016Borisova}. Since CIV is a resonant line, the extended  CIV emission in both QSOs could also arise due to the resonant scattering by the same medium scattering Ly\al~with non-detection in the extended non-resonant HeII line. HeII does not have an extended structure in QSO2, but QSO1 has a faint extended emission signature in miniJPAS images. Therefore, the extended emission for HeII in QSO1 may be powered by a different scenario. 

The double-peaked line profile in Ly\al~spectrum of QSOs at high redshift is common due to the resonant nature of Ly\al~\citep{Dijkstra2006, 2007Tapken, Yang2014, Cai2017}.
The Ly\al~line profile shows asymmetric due to the bulk motion of neutral hydrogen and Ly\al~photons escape in a double-peaked line due to the high optical depth at the line center and it is absorbed and re-emitted in another direction \citep{2021Sanderson, 2007Dijkstra, 2018Matthee}. Consequently, Ly\al~photons diffuse spatially and also in frequency space. The double-peak profile in the Ly\al~emission line seen in the SDSS spectra of both QSO1 and QSO2 indicates that the Ly\al~emission is powered by resonant scattering. Ly\al~line of both QSOs are redshifted with respect to other emission lines ($\Delta$v = 318.79 km/s for QSO1 and $\Delta$v = 1052.39 km/s for QSO2) suggests that the Ly\al~photons scatter through large-scale outflows \citep{2006Verhamme, 2011Steidel}. 
The galactic outflows (inflows) would also lead to an enhanced red (blue) peak in the Ly\al~profile \citep{2002Zheng, 2006Verhamme, 2007Dijkstra, 2011Laursen, 2018Weinberger}. Moreover, IGM resonant scattering causes the diminishing of the blue peak. The simulation results from \cite{2011Laursen} indicate that at z$=$3.3, IGM transmission in the blue side of Ly\al~line is 80\%. The Ly\al~profile of both QSOs shows an enhanced red peak with a diminished blue peak. Such spectral features indicate gas outflows. Therefore, galactic outflows play a role in powering Ly\al~emission in both QSOs and show redshifted Ly\al~profile with enhanced red Ly\al~peak.
\subsection{Origin of the double-core Ly\al~morphology}
Here, we briefly outline the physics underlying the different scenarios that may explain double-core morphology in Ly\al~emission. 

Double-cored Ly\al~has been seen in some high-$z$ radio galaxies, where it appears to be related to jet interactions \citep[e.g.,][]{2001Overzier}, but neither QSO1 or QSO2 are radio-loud galaxies. Gravitational lensing is one of the scenarios that can explain the unusual morphology in QSOs \citep{1979Walsh, 2019Meyer}.
However, the single-core QSO continuum component in miniJPAS BB (e.g., rSDSS) and NB images (e.g., J0490 in QSO1 and J0460 in QSO2) lacks evidence for lensed QSOs or the corresponding lensing galaxy, or even  binary QSOs. The single-point structure in the HST F606W and F814W images of QSO1 also disfavor the QSO lensing hypothesis. 
Ionization echoes are often seen from type 2 AGN/QSO, as the echoes were enlightened by the previously active but now inactive SMBH. AGN ionization echoes are fossil records of the rapid shutdown of luminous QSOs, are uncovered in low-$z$ ($z=0.05$ to 0.35) characterized by high-ionization gas extending more than 10kpc from AGN and show strong [OIII]\lam~5007 emission powered by type 2 AGN \citep{2010Schawinski, 2012Keel, 2015Keel, 2013Schweizer}. The two QSOs reported here are all type 1 QSOs. Since these two QSOs have broad emission lines in their spectra, and AGN counterparts, we can rule out the AGN ionization echoes hypothesis.

The morphology of Ly\al~emission due to outflows strongly depends on the orientation of the outflow with respect to the line of sight. Outflows emerging from QSOs show double-lobed (bipolar) structure in their morphology if viewed perpendicular to the cone axis \citep{2005Weidinger, 2016Borisova, 2021Sanderson}. The morphology of the miniJPAS Ly\al~image and line profile from SDSS spectra support the hypothesis that outflows in these two QSOs are contributing to the double-core Ly\al~morphology in the miniJPAS image. Alternatively, it is possible that high central HI column density or dust obscuration may be the reason for the asymmetric double-core structure in QSO2. \cite{2019Erb} explained the offset by a significant variation of the neutral hydrogen column density across the object. \cite{2021Chen} identified 2 Ly\al~ nebulae spatially offset from the associated star-forming regions, suggesting large spatial fluctuations in the gas properties \citep[see also][]{Claeyssens2022}. \cite{2016Borisova} found no evidence for "bipolar ionization cone illumination" patterns in their study of Ly\al~ nebulae around quasars, except for one quasar (see their Section. 4.2).

\section{Conclusions}
We uncovered the double-cored morphology in the Ly\al~NB image of two QSOs: SDSS-J141935.58+525710.7, and SDSS-J141813.40+525240.0 at $z=3.218$ and $z=3.287$, respectively, during the search for extended Ly\al~around high-z QSOs in the miniJPAS survey. Our results are summarized as follows:
\begin{itemize}
    \item The separations of the two Ly\al~cores are 11.07 $\pm$ 2.26 kpc (1.47 $\pm$ 0.3$^{\prime\prime}$) and 9.73 $\pm$ 1.55 kpc (1.31 $\pm$ 0.21$^{\prime\prime}$) with Ly$\alpha$~line luminosities of $\sim$ 3.35 $\times 10^{44}$ erg s $^{-1} $ and $\sim$ 6.99 $\times$ 10$^{44}$ erg s $^{-1}$ for QSO1 and QSO2, respectively.
    \item The Ly$\alpha$ luminosity places these Ly$\alpha$ nebulae at the high luminosity end in the luminosity-size diagram of the few previous detections of ELANs, suggesting that deeper observations might reveal the large-scale ELAN structure of these two QSOs.
    \item The spatially distributed double-core morphology in Ly\al~images might be due to the scattering of Ly\al~photons through galactic outflows in bi-conical structure. 
    \item Both QSOs show spatially extended strong CIV emission ($>$ 30 kpc) in miniJPAS images, suggesting that halo gas in both QSOs are metal-rich and powered by collisional excitation by shocks or photoionization by AGN. The presence of faint extended HeII emission in QSO1 indicates an additional contribution from collisional ionization due to shocks.
    \item This pilot study demonstrates the capability of J-PAS/miniJPAS to identify a large number of Ly\al~nebulae candidates by looking at the morphology in the Ly\al~NB filter (and other CIV and HeII NB filters).
    
\end{itemize}
It is quite unique to discover such a double-core Ly\al~morphology from relatively shallow NB imaging surveys such as miniJPAS which covers only 1 deg$^2$ of the sky. The curious Ly\al~ morphology of these QSOs may shed new light on the origin of these types of nebulae. The future J-PAS (will cover 8000 deg$^2$) may well have the capacity to discover many such objects ($\sim1$ per deg$^2$), warranting this pilot project.
With the current miniJPAS observation, it is very hard to draw a conclusion about the primary driving mechanism for the origin of Ly\al~emission. We are planning for deep and wide field spectroscopic observations to make a more definitive statement about the kinematics, ionization status, metallicity, and driving mechanism as well as to trace the large-scale low SB level diffuse region of Ly\al~emission. In a forthcoming paper, we discuss more cases of extended Ly\al~emission around high-z QSOs detected in the miniJPAS survey. These studies demonstrate the capabilities of contiguous NB imaging like JPAS survey to study the high-z Ly\al~nebulae. 

\section*{Acknowledgements}

We would like to thank the referee for their very helpful comments and suggestions. Z.Y.Z. acknowledges support by the National Science Foundation of China (12022303), the China-Chile Joint Research Fund (CCJRF No. 1906), and the CAS Pioneer Hundred Talents Program. RPT thanks the CAS President’s International Fellowship Initiative (PIFI) (Grant No. E085201009) for supporting this work. We acknowledge the science research grants from the China Manned Space Project with No. CMS-CSST-2021-A04 and CMS-CSST-2021-A07.

ACS acknowledges funding from the Conselho Nacional de Desenvolvimento
Científico e Tecnológico (CNPq) and the Rio Grande do Sul Research Foundation (FAPERGS) through grants CNPq-11153/2018-6, CNPq-314301/2021-6 and FAPERGS/CAPES 19/2551-0000696-9 and the Chinese Academy of Sciences (CAS) President's International Fellowship Initiative (PIFI) through grant E085201009.

R.P.T and Z.Y.Z thanks Shuairu Zhu, Fang-Ting Yuan, Ruqiu Lin, and Xiang Ji for their useful discussion during the preparations of the manuscript. R. P. T acknowledges Carolina Queiroz for sharing information about the JPAS database.

This paper has gone through the internal review by the J-PAS collaboration.
Based on observations made with the JST/T250 telescope and JP Cam at the Observatorio Astrofísico de Javalambre (OAJ), in Teruel, owned, managed, and operated by the Centro de Estudios de Física  del Cosmos de Aragón (CEFCA). We acknowledge the OAJ Data Processing and Archiving Unit (UPAD) for reducing and calibrating the OAJ data used in this work. 

Funding for the J-PAS Project has been provided by the Governments of Spain and Aragón through the Fondo de Inversión de Teruel, European FEDER funding and the Spanish Ministry of Science, Innovation  and  Universities,  and  by  the  Brazilian  agencies  FINEP,  FAPESP,  FAPERJ and  by  the National Observatory of Brazil. Additional funding was also provided by the Tartu Observatory and by the J-PAS Chinese Astronomical Consortium.

L.A.D.G., C.K., and R.G.D., acknowledge financial support from the State Agency for Research of the Spanish MCIU through the "Center of Excellence Severo Ochoa" award to the Instituto de Astrofísica de Andalucía (SEV-2017-0709), and to PID2019-109067-GB100.



\bibliography{ref}{}
\bibliographystyle{aa}



\end{document}